\documentstyle[12pt]{article}
\begin{document}
\setlength{\unitlength}{1mm}
\newcommand{\te}{\theta}
\renewcommand{\thefootnote}{\fnsymbol{footnote}}
\newcommand{\gapproxeq}{\lower .7ex\hbox{$\;\stackrel{\textstyle >}{\sim}\;$}}
\newcommand{\lapproxeq}{\lower .7ex\hbox{$\;\stackrel{\textstyle <}{\sim}\;$}}
\newcommand{\bee}{\begin{equation}}
\newcommand{\ene}{\end{equation}}
\newcommand{\bea}{\begin{eqnarray}}
\newcommand{\ena}{\end{eqnarray}}
\newcommand{\dx}{d(x)}
\newcommand{\ux}{u(x)}
\newcommand{\sx}{s(x)}
\newcommand{\adx}{\bar{d}(x)}
\newcommand{\aux}{\bar{u}(x)}
\newcommand{\asx}{\bar{s}(x)}
\newcommand{\Dgx}{\Delta G(x)}
\newcommand{\Ddx}{\Delta d(x)}
\newcommand{\Dux}{\Delta u(x)}
\newcommand{\Dsx}{\Delta s(x)}
\newcommand{\Dadx}{\Delta \bar{d}(x)}
\newcommand{\Daux}{\Delta \bar{u}(x)}
\newcommand{\Dasx}{\Delta \bar{s}(x)}
\newcommand{\tra}{\triangle\theta_}
\newcommand{\uupv}{u^{\uparrow}_{val}}
\newcommand{\udwv}{u^{\downarrow}_{val}}
\newcommand{\dupv}{d^{\uparrow}_{val}}
\newcommand{\ddwv}{d^{\downarrow}_{val}}
\newcommand{\auupv}{\bar{u}^{\uparrow}_{val}}
\newcommand{\audwv}{\bar{u}^{\downarrow}_{val}}
\newcommand{\adupv}{\bar{d}^{\uparrow}_{val}}
\newcommand{\addwv}{\bar{d}^{\downarrow}_{val}}
\newcommand{\uupx}{u^{\uparrow}(x)}
\newcommand{\udwx}{u^{\downarrow}(x)}
\newcommand{\dupx}{d^{\uparrow}(x)}
\newcommand{\ddwx}{d^{\downarrow}(x)}
\newcommand{\auupx}{\bar{u}^{\uparrow}(x)}
\newcommand{\audwx}{\bar{u}^{\downarrow}(x)}
\newcommand{\adupx}{\bar{d}^{\uparrow}(x)}
\newcommand{\addwx}{\bar{d}^{\downarrow}(x)}
\newcommand{\fnx}{{F_{2}^{n}}(x)}
\newcommand{\fpx}{{F_{2}^{p}}(x)}
\newcommand{\gpx}{{g_{1}^{p}}(x)}
\newcommand{\gnx}{{g_{1}^{n}}(x)}
\newcommand{\gdx}{{g_{1}^{d}}(x)}

{{\hfill \small $\begin{array}{r}\mbox{
Universit\'a di Napoli Preprint DSF-16/96 (april 1996)} \\
\mbox{hep-ph/9604230}\end{array}$}}

\begin{center}
{\Large\bf Quantum Statistical Parton Distributions and the Spin Crisis}
\end{center}
\bigskip\bigskip

\begin{center}
{{\bf F. Buccella}, {\bf G. Miele} and {\bf N. Tancredi}}
\end{center}

\bigskip

\noindent
{\it Dipartimento di Scienze Fisiche, Universit\`a di Napoli 
''Federico II'', and INFN Sezione di Napoli,
Mostra D'Oltremare Pad. 20, I--80125 Napoli, Italy}

\bigskip\bigskip\bigskip

\begin{abstract}
Quantum statistical distributions for the partons provide a fair
description of deep inelastic scattering data at $Q^2 = 3$ and
$10~(GeV/c)^2$. The study of the polarized structure functions seems
to suggest an alternative possible solution of the {\it spin crisis}
based on the Pauli principle. In this scheme, in fact, the defects of
the Gottfried sum rule and Ellis--Jaffe sum rule for proton, result
strongly connected. This possibility finds particular evidence from
the phenomenological observation that the relation $\Delta u = 2
\tilde{F} + u - d -1$ seems well satisfied by parton distributions. 
\end{abstract}

\baselineskip22pt
\newpage

\section{Introduction}

The experimental results on deep inelastic scattering (DIS) are always
an inexhaustible source for deeper insight in the nucleon structure.
Among them, the violation of well established sum rules represent the
relevant starting point to unveil the mechanisms which rule the parton
physics. 

This consideration has inspired a recent series of papers, which
starting from an old idea of Ni\'egawa and Sasaki \cite{nisa} and 
Field and Feynman \cite{Fey}, have
focused the role played by Pauli exclusion principle on the
quark/parton distributions inside nucleons \cite{bs}--\cite{bs2}. In
this framework, by virtue of the fermionic statistics, the violation
of the Gottfried sum rule \cite{Gott} and the Ellis--Jaffe sum rule for
the proton \cite{EJ} are related. Interestingly, this connection has
also been observed in a more general framework by using standard
parameterizations for parton distributions \cite{bpss}--\cite{Pisanti}. 

Moreover, the statistical inspired approach to the parton
distributions has suggested a parameterization of quark/gluon
distributions in terms of Fermi--Dirac/Bose--Einstein statistical
functions. Remarkably, one obtains a satisfactory description of the
experimental data of DIS in terms of few free parameters
\cite{bbmmst}. The analysis has then been improved \cite{bmmt} by
adding an extra contribution to the {\it statistical} term of parton
distributions, dominating in the small $x$ region, the so called {\it
liquid} component, with Pomeron--like quantum numbers. 

There are several motivations for further work. The first one is the
necessity of considering the polarized data with deuteron target at
SLAC (E143) \cite{g1dE143} and at CERN (SMC) \cite{g1dSMC}. These
measurements, differently from the data obtained on neutron target at
SLAC (E142) \cite{E142}, seem to be consistent with Bjorken (Bj) sum
rule \cite{Bj} when QCD corrections are taken into account.\\ The
second reason is to consider sets of data corresponding to the same
$Q^2$. In this respect, we choose for $Q^2$ the values $3$ and
$10~(GeV/c)^{2}$, at which E143 and SMC are performed. Indeed, in the
previous analysis the data used for $xF_3$ and on $\fpx$ ($\fnx$) were
taken at slightly different $Q^2$ ($3$ and $4~(GeV/c)^2$), and
moreover one tried to fit the experimental results of $x\gpx$
corresponding to $Q^2=3~(GeV/c)^{2}$ and $10~(GeV/c)^{2}$ by means of
a single function, since they did not differ too much. Finally, for
$x\gnx$ the data were at $Q^2 = 2 ~(GeV/c)^{2}$.\\ The third reason is
to explore the possibility of gluons and/or strange quark
polarization, which have been advocated  to explain the defect in the
Ellis--Jaffe (EJ) sum rule \cite{EJ} for the proton in a framework
consistent with the Bj sum rule. 

Concerning the very important issue of testing the Bjorken sum rule,
we want to stress the difficulty to evaluate the small $x$
contribution, where the measured cross section has to be divided by
$x$. In this respect, it would be helpful to know the small $x$
behaviour of $\gpx - \gnx$. In the approach based on statistical
functions this behaviour is given by the function $f(x)$ \cite{bmmt},
which is also related to $\fpx - \fnx$ and $xF_{3}(x)$, occurring in
the Gottfried and Gross Llewellyn--Smith sum rule \cite{GLl},
respectively. Since these quantities are measured more precisely than
the $g_1$'s, $f(x)$ is practically determined by the unpolarized
structure functions. Once $f(x)$ and the parameter $\bar{x}$ are
established \cite{bbmmst,bmmt}, the polarized structure function 
$\gpx-\gnx$ depends only on the {\it thermodynamical potentials}
\cite{bbmmst,bmmt} of quarks with definite spin. In practice, we have
only one free parameter for each quark, since the sum $p^{\uparrow}(x)
+ p^{\downarrow}(x)$ is fixed by the unpolarized structure functions.
In this way, we are able to obtain from the data the values of the
parameters and an evaluation of the Bj sum rule less dependent on the
large errors of the experimental data at small $x$. 

We consider the data at $Q^2=3~(GeV/c)^{2}$ and $10~(GeV/c)^{2}$
as independent, without relating them by the evolution equations
\cite{evolv}. In principle, one should test whether, starting from
quantum statistical distributions at a given $Q^2$ one gets the
same form for the distributions at higher $Q^2$, apart from some changes
for the parameters. Practically, one has enough flexibility in our 
parameterization to reproduce the changes induced by the evolution equations, 
which bring to a narrowing of the parton distributions and of the structure 
functions.\\
The possibility that the evolution equations should be modified to
take into account the statistical properties of partons has been studied,
within the approximation to consider for parton momenta the longitudinal 
degree of freedom only \cite{mmm}. The very important role of transverse 
degrees of freedom, however, demands that they should be taken into
account in a model aiming to be realistic. 

The paper is organized as follows, in the second section we give a
brief review of the role of Pauli principle to explain some features
of deep inelastic scattering (DIS) data. Section three is devoted to
describe the experimental results used for the numerical analysis. The
results of the fit performed are shown in section four and finally, in
section five we give our conclusions. 

\section{The Pauli principle and the structure functions}

In the usual description for the deep inelastic phenomena, the
quark/partons inside the nucleons are seen, in the infinitum momentum
frame, as an ensemble of free particles which incoherently interact
with the electroweak probe. In this scenario, no statistical effects
are considered since they would depend on the overlapping of the quark
wave functions. 

A role of Pauli principle has been advocated to understand the defect
in the Gottfried sum rule \cite{Gott}, and in the Ellis--Jaffe sum
rule for the proton \cite{EJ}.\\ According to the Gottfried sum rule,
one gets 
\bee
I_{G} \equiv \int_{0}^{1} { dx\over x}\left[\fpx - \fnx \right] = { 
1\over 3} +{ 2 \over 3} \int_{0}^{1}dx~\left[ \bar{u}(x) - \bar{d}(x)
\right] = { 1 \over 3}~~~,
\label{2.1}
\ene
where the last equality holds in the limit of $SU(2)_{I}$--symmetric
sea ($\bar{d} = \bar{u}$). The experimental measurement of $I_G$, by
NMC \cite{f2p-3g-2}, gives 
\bee
I_{G} = 0.235 \pm 0.026~~~,
\label{2.2}
\ene
which implies, assuming the validity of Adler sum rule \cite{Adler}, a
strong violation of the isospin invariance of sea quarks 
\bee
\bar{d} - \bar{u} = 0.15 \pm 0.04 ~~~,~~~~~~~~~~~ u - d = 0.85 \pm 0.04~~~.
\label{2.3}
\ene
As far as the Ellis--Jaffe sum rule is concerned \cite{EJ}, for the
proton it reads \cite{Altarelli} 
\bea
\Gamma^p_1  &\equiv& \int_{0}^{1} dx~\gpx 
=  \int_{0}^{1} dx~\left[{2 \over 9} (\Dux + \Daux)
+ { 1 \over 18} (\Ddx + \Dadx \right.\nonumber\\
&+& \left. \Dsx + \Dasx) - { \alpha_s(Q^2)
\over 6 \pi} \Dgx \right]
=  {\tilde{F}(Q^2) \over 2} - { \tilde{D}(Q^2) \over 18}~~~,
\label{2.4}
\ena
where the parameters $\tilde{F}(Q^2)$ and $\tilde{D}(Q^2)$ are given,
up to the third order in $\alpha_s(Q^2)$, by the following expressions
\cite{Kodaira} 
\bea
\tilde{F}(Q^2) & = & F - { \alpha_s(Q^2) \over 5 \pi}
\left( 3~F + { 2 \over 3} D \right) - \left({ \alpha_s(Q^2) \over \pi}
\right)^2 (2.092~F+0.496~D)\nonumber\\
& - & \left({ \alpha_s(Q^2) \over \pi} \right)^3 4.044~(2~F+D)~~~,
\label{2.4a}\\
\tilde{D}(Q^2) & = & D - { \alpha_s(Q^2) \over 5 \pi}
\left( 2~F + { 13 \over 3} D \right) - \left({ \alpha_s(Q^2) \over \pi}
\right)^2 (1.488~F+3.084~D)\nonumber\\
& - & \left({ \alpha_s(Q^2) \over \pi} \right)^3 4.044~(3~F+4~D)~~~.
\label{2.4b}
\ena
Note that the last equality of Eq. (\ref{2.4}) is obtained by
neglecting the polarization of strange quarks and gluons. By using the
values of $F$ and $D$ determined in Ref. \cite{FD} 
\bee
F=0.46\pm 0.01~~~,~~~~~~~~~D=0.79 \pm 0.01~~~,
\label{2.5}
\ene
one gets from (\ref{2.4}): $\Gamma^p_1 = .161 \pm .007$ at $Q^2 = 3
~(GeV/c)^2$ ($\alpha_s(3 ~(GeV/c)^2)=0.35 \pm 0.05$), and $\Gamma^p_1
= 0.169 \pm .005$ at $Q^2 = 10 ~(GeV/c)^2$ ($\alpha_s(10
~(GeV/c)^2)=.27\pm.02$). These two quantities have been measured by
E143 \cite{g1pE143} and SMC \cite{g1pSMC} respectively, and result to
be $\Gamma^p_1 = 0.127 \pm 0.004 \pm 0.010$ at $Q^2 = 3 ~(GeV/c)^2$
and $\Gamma^p_1 = 0.136 \pm 0.011 \pm 0.011$ at $Q^2 = 10 ~(GeV/c)^2$.
The disagreement between the theoretical predictions and the
experimental data, by virtue of Eq.(\ref{2.4}), is typically explained
in terms of a relevant polarization of strange quarks and/or gluons.
As far as the strange quarks are concerned, since their polarization
cannot exceed the less divergent part of the unpolarized distribution
\cite{PS,PSR} they cannot explain the EJ results, whereas the gluons
due to the suppression factor $\alpha_s/6 \pi$ need very strong
polarization to be effective. This leads to a very unnatural scenario,
in which the relevant gluon polarization has to be compensated by a
strong angular orbital momentum of the partons. Furthermore, if we
admit such a large anomaly contribution to explain the EJ sum rule on
proton it would imply the dominance at low $x$ of $\Delta G(x)$ over
the other parton polarizations, and thus one would expect for $\gpx$
and $\gnx$ the same behaviour at low $x$ by virtue of the isoscalar
nature of gluons. This would contradict the trend, shown by the
experimental data of SMC on proton \cite{g1pSMC} and deuteron
\cite{g1dSMC}, of a different behaviour at low $x$ for $\gpx$ and
$\gdx$. 

Interestingly, the defects in the two sum rules may be easily
explained in terms of the fermionic nature of quarks. In fact, the
violation of the Gottfried sum rule can be understood by following the
idea of Ni\'egawa and Sasaki \cite{nisa} and 
Field and Feynman \cite{Fey}, that Pauli principle
disfavours the production of $u \bar{u}$ pairs in the proton with
respect to $d \bar{d}$, since it contains two valence $u$ quarks and
only one $d$. For the same reason, the violation of Ellis--Jaffe sum
rule for proton may be explained by observing that at high $x$ the
$\uupx$ is the dominating parton distribution in the proton and again
Pauli principle would disfavour contributions from the sea to
$u^{\uparrow}$ . 

Information on the parton distributions can be obtained from their
first momenta. Indeed at $Q^2 =0$, the axial couplings of the baryon
octet are fairly described in terms of the valence quarks 
\bee
\uupv  = 1+F~~~,~~~~~~~~~\udwv  = 1-F~~~, 
\label{2.6}
\ene
\bee
\dupv  = {1+F-D \over 2}~~~,~~~~~~~\ddwv  = {1-F+D \over 2}~~~. 
\label{2.7}
\ene
Thus, by using the experimental values for $F$ and $D$ \cite{FD}
\bee
\uupv \simeq { 3 \over 2} \simeq \udwv + \dupv + \ddwv~~~.
\label{2.8}
\ene
This result on the abundances of the {\it valence} quarks, suggests to
assume a similar relation for the parton distributions \cite{bs}. The
leading idea, the so--called {\it shape--abundance} correlation, is
that to more abundant partons correspond broader distribution
functions, as stated from the Pauli principle. This assumption, which
is a key property of the Fermi--Dirac distribution functions, can be
tested by comparing the shapes of $\uupx$ with the ones of the other
partons. 

Indeed, one can observe that the behaviour at high $x$ of $\fnx/\fpx$
\cite{f2p-3g-2}, which experimentally tends to $\approx 1/4$, indicates the
dominance of the $u$ quarks, and in addition, the study of polarized
$\gpx$ shows that at high $x$ the partons with spin parallel to the
proton dominate. In conclusion, at high $x$, the most abundant parton,
$u^{\uparrow}$, dominates according to the already mentioned
correlation shape--abundance. 

As a consequence of this analysis, the Pauli principle seems to play a
fundamental role in the DIS phenomena. Thus, in this framework it is
natural to assume Fermi--Dirac distributions in the variable $x$ for
the quark partons \cite{bbmmst,bmmt} 
\bee
p(x) = f(x) \left[\exp\left({ x - \tilde{x}(p) \over \bar{x}} \right) + 1
\right]^{-1}~~~,
\label{2.13}
\ene
where $f(x)$ is a weight function, $\bar{x}$ plays the role of the
temperature and $\tilde{x}(p)$ is the {\it thermodynamical potential}
of the parton $p$, identified  by its flavour and spin direction.\\
Analogously, for the gluons one should have 
\bee
G^{\uparrow(\downarrow)}(x) = {8 \over 3} f(x) 
\left[\exp\left({ x - \tilde{x}(G^{\uparrow(\downarrow)}) \over \bar{x}} 
\right) - 1 \right]^{-1}~~~.
\label{2.14}
\ene
The factor $8/3$ in (\ref{2.14}) is just the ratio of the colour
degeneracies for gluons and quarks.\\ As far as the weight function
$f(x)$ is concerned, it contains the information on the density of
states for quarks and gluons  inside the nucleon in the $P_z = \infty$
frame of reference. Nevertheless, it has to vanish at $x=1$, and one
should recover the usual parameterization for parton distributions
when the statistical effects are negligible, hence we assume for it
the usual form 
\bee
f(x)= A~ x^{\alpha} (1 -x)^{\beta}~~~,
\label{2.15}
\ene
where $A$, $\alpha$ and $\beta$ are free parameters. Indeed, the
expressions (\ref{2.13}) and (\ref{2.14}), assumed for quarks and
gluons, are not able to recover the typical behaviour of the
distribution functions at low $x$. In fact, the most divergent part of
distributions is expected on general grounds to be equal for the
different partons, at least in the limit of flavour symmetry, 
and hence it does not contribute to the sum rules, but
affects other DIS observables. In order to disentangle, this divergent
term from the total distribution function we add a {\it liquid}
unpolarized component for the light quark--partons ($u$, $d$ and their
antiparticles) \cite{bmmt} 
\bee
f_L(x) = (A_{L}/2)~x^{\alpha_{L}} (1-x)^{\beta_L}~~~,
\label{2.16}
\ene
and the same $x$ dependence, but with a different normalization for 
$s$ and $\bar{s}$.

\section{The experimental data}

In a previous paper \cite{bmmt} the above analysis to determine the
parton distributions in the {\it statistical inspired approach}, was
performed by choosing sets of experimental data obtained at different
$Q^2$. In particular, as already stated in the introduction, the
experimental results for $xF_3$ \cite{xf3-3g}, and $\fpx$, $\fnx$
\cite{f2p-3g-2} were taken at slightly different $Q^2$, $3$ and
$4~(GeV/c)^2$, respectively. For the polarized function $\gpx$ one
also tried to describe both data at $Q^2=3$ \cite{g1pE143} and
$10~(GeV/c)^2$ \cite{g1pSMC} with the same function. Nevertheless,
even in this approximation, due to the smooth dependence of the DIS
observables on the $Q^2$ in the range $3-10~(GeV/c)^2$, one found a
good agreement between the experimental results and the theoretical
predictions \cite{bmmt}. 

In the following analysis we will use sets of data which correspond to
the same values of $Q^2$. In particular we will perform our fit for
the values of $Q^2 = 3~(GeV/c)^2$, and $10~(GeV/c)^2$, corresponding
to the measurements of $\gpx$ and $\gdx$ by E143 \cite{g1dE143,g1pE143},
and SMC \cite{g1dSMC,g1pSMC}, respectively.
For the same values of $Q^2$, we will also analyze the unpolarized
data on $xF_3(x)$, $\fpx$ and $\fnx$. 

As long as $xF_3(x)$ is concerned, CCFR \cite{xf3-3g} gives directly
the data at $Q^2=3~(GeV/c)^2$, whereas to get the results at
$Q^2=10~(GeV/c)^2$ we use a linear interpolation in $\log{Q^2}$. For
the ratio $\fnx/\fpx$, we start from the data at $Q^2=4~(GeV/c)^2$
reported in Ref. \cite{f2p-3g-2} and get the values at $Q^2=3$ and
$10~(GeV/c)^2$, by means of the formula 
\bee
{ F^n_2(x,Q^2) \over F^p_2(x,Q^2) } = { F^n_2(x,4~(GeV/c)^2) 
\over F^p_2(x,4~(GeV/c)^2) } + B(x) \log\left[{Q^2 \over 4
~(GeV/c)^2}\right]~~~,
\label{3.1}
\ene
written in Ref. \cite{f2n/f2p}, where $B(x)$ is also given.

For this parameterization the systematic errors are evaluated
according to Ref. \cite{f2p-3g-2}, whereas the statistical ones are
taken from the measurements at $Q^2=4~(GeV/c)^2$ \cite{f2p-3g-2}.
Finally, the data on $\fpx$ are taken from Hera, SLAC and NMC 
\cite{f2p-3g-2}, \cite{f2p-3g-1},  \cite{f2p-3g-3}. 

We will also take into account the NA51 result for the Drell--Yan
processes on proton and deuteron targets \cite{NA51}, which depends
mainly on the ratio $\bar{u}(.18)/\bar{d}(.18)$ and provides the
measurement for this ratio $0.51 \pm 0.04 \pm 0.05$, and as far as the
gluons are concerned, the unpolarized theoretical distribution, which
we impose to carry the fraction of proton momentum not carried by the
quark partons, will be compared with the experimental results reported
in \cite{gluoni1}--\cite{gluoni3}. 

\section{Comparison with experiment}

As in Ref. \cite{bmmt} we take for quark/parton distributions the following
expression
\bee
p(x) = f_L(x)+ 
A x^{\alpha} (1-x)^{\beta} \left[ \exp\left({x -\tilde{x}(p) \over
\bar{x}}\right) + 1 \right]^{-1}~~~.
\label{4.1}
\ene
The parameters $A_{L}$, $\alpha_L$ and $\beta_L$ occurring in the
definition of $f_L(x)$ (see Eq. (\ref{2.16})), define the low $x$
behaviour of parton distributions, independently of the parton specie,
the so--called {\it liquid} term, whereas $A$, $\alpha$ and $\beta$
fix the weight function in the {\it gas} term. Finally, the
statistical functions, which depend on the parton are simply defined
by an {\it universal temperature} parameter, $\bar{x}$, and by the
{\it thermodynamical potentials} $\tilde{x}(p)$. Note that, $p$ stands
for partons with given flavour and polarization. For the strange
quarks, we take for their unpolarized structure functions the
following expression 
\bee
s(x) = \bar{s}(x) = A_s {\bar{u}(x) + \bar{d}(x) \over 2}~~~,
\label{4.2}
\ene
with $A_s=.475$ according to Ref. \cite{As}.
Here, we also allow for polarized gluon distributions
\bee
G^{\uparrow(\downarrow)}(x) = {8 \over 3} 
A x^{\alpha} (1-x)^{\beta} \left[ \exp\left({x -\tilde{x}
(G^{\uparrow(\downarrow)}) \over \bar{x}}\right) - 1\right]^{-1}~~~.
\label{4.3}
\ene
Should we consider also the {\it liquid} component for gluons, 
the momentum fraction carried by the gas component should  be
smaller, giving rise to smaller {\it thermodynamical} potentials
$\tilde{x}(G^{\uparrow(\downarrow)})$. However, since only
$\Delta G(x)$ appears in the structure functions, the error
implied by neglecting the {\it liquid} gluon component,
which is expected to carry only a small fraction of the total 
momentum, it is not relevant for the issue of deciding 
is a Bose/Einstein gluon contribution may solve the spin crisis.\\
Of course, to get a fair description of the gluon distributions, one
should have also a {\it liquid} gluon component. Since, the low $x$
gluon data are at $Q^2=20~(GeV/c)^{2}$ \cite{gluoni1}--\cite{gluoni3},
while we try to reproduce the structure functions at $Q^2=3$ and
$10~(GeV/c)^{2}$, where $\gpx$ and $\gdx$ are measured, a description
of $G(x)$ goes beyond the purpose of this work.

The large negative values found in Ref. \cite{bmmt} for
$\tilde{x}(\bar{u})$ and $\tilde{x}(\bar{d})$ imply negligible effects
of Pauli principle for the $\bar{q}$'s (the levels are very little
occupied) and when we can neglect $1$ in the denominator of
(\ref{4.1}) the distribution become Boltzman--like and with the same
$x$ dependence. Thus, it is a good approximation within our general
approach to put 
\begin{eqnarray}
\Daux & \simeq & k_u \left[\bar{u}(x)- 2 f_L(x)\right]~~~,\label{4.4}\\
\Dadx & \simeq & k_d \left[\bar{d}(x)- 2 f_L(x)\right]~~~,\label{4.5}\\
\Dsx + \Dasx & \simeq & k_s \left[s(x) + \bar{s}(x) - 1.9 f_L(x)
\right]~~~,\label{4.6}
\end{eqnarray}
with $|k_u|,|k_d|,|k_s|\leq 1$. We tried, in our approach, several
options to see how much the different solutions proposed for spin
crisis are supported by data. In Tables I.a and I.b, we report for
four different options, the values of normalized $\chi^2$, $\Delta
q_i$ and $\Delta \bar{q}_i$ (with $i=u,d,s$) and $\Delta G_i$, and
moreover the corresponding evaluations of the EJ and Bj sum rules
(with in brackets the expected values up to the third order in QCD
corrections)for $Q^2=3$ and $10~(GeV/c)^2$, respectively. 

The following remarks are then in turn. 

\noindent
a) To summarize the results of Tables I.a and I.b one can observe that 
at $Q^2=3~(GeV/c)^2$
\begin{itemize}
\item $.640\leq \Delta u + \Delta \bar{u} \leq .674$ to be compared
with the QCD prediction $.782$ up to $\alpha_s^3$;
\item $-.300 \leq \Delta d + \Delta \bar{d} \leq -.260$ versus 
the QCD prediction $-.238$;
\item finally, for the Bj sum rule one gets 
$.155 \leq ( \Delta u + \Delta \bar{u} - \Delta d - \Delta \bar{d})/6
\leq .157$ to be compared with $.170$ .
\end{itemize}
The same quantities for $Q^2=10~(GeV/c)^2$ result to be
\begin{itemize}
\item $.670\leq \Delta u + \Delta \bar{u} \leq .730$ to be compared
with $.826$ from QCD;
\item $-.290 \leq \Delta d + \Delta \bar{d} \leq -.183$ versus $-.268$;
\item $.156 \leq ( \Delta u + \Delta \bar{u} - \Delta d - \Delta \bar{d})/6
\leq .160$ to be compared with $.183$ .
\end{itemize}
Thus, in both cases $\Delta u + \Delta \bar{u}$ is smaller than the
QCD result, while only at $Q^2 = 3~(GeV/c)^2$ we have a $\Delta d +
\Delta \bar{d}$ more negative than the QCD value. It is somehow
surprising the result found for the Bj sum rule at $Q^2
= 10~(GeV/c)^2$. We find a defect, while from its own data SMC gets an
excess (consistent within the errors with the predicted value). The
origin of this discrepancy can be understood as a consequence of our
assumption of a common behaviour at low $x$ of the parton component
contributing to the sum rules. In this way, in fact, we constrain the
contribution to the polarized sum rule to have the same behaviour in
the $x \rightarrow 0$ limit of the unpolarized ones. The value found
for $\bar{x} \gapproxeq .2$ gives a very smooth dependence at small
$x$ for the ratio of the different parton distributions. In fact, in
this $x$ region we are unable to reproduce the measurements of
$x\gpx$, and the fast decrease to negative values of $x\gdx$. The
Bose--Einstein form chosen for the gluons enhances their contributions
at low $x$, but cannot account for a different behaviour of $x\gpx$
and $x\gnx$ in that limit for its isoscalar nature. In fact our fit,
with almost half the contribution to the $\chi^2$ coming from deuteron
data despite their large errors, is unable to reproduce these typical
features of SMC data. 

\noindent
b) The possibility to have $\Delta \bar{d} \neq 0$ leads to very
unphysical situations, namely, very negative $\Delta d$, partially
compensated by positive $\Delta \bar{d}$ at $Q^2=3~(GeV/c)^2$, and
small positive $\Delta d$ with larger negative $\Delta \bar{d}$ at
$Q^2=10~(GeV/c)^2$. These results come from the difficulty of the fit
procedure to disentangle in $\Delta \bar{d}+ \Delta d$ the single two
contributions, which in the previous paper led us to assume $\Delta
\bar{d}=0$. Since, when we allow for $\Delta \bar{d} \neq 0$, the sum
$\Delta \bar{d}+ \Delta d$, is more negative than for $\Delta
\bar{d}=0$ at $Q^2=3~(GeV/c)^2$ and less negative at
$Q^2=10~(GeV/c)^2$, we think it is better to assume also in the
present analysis $\Delta \bar{d}=0$. The sum $\Delta u + \Delta
\bar{u}$ does not change so much when we take $\Delta \bar{u} \neq 0$.
Finally, one can observe that the strange quark polarization $(\Delta
s + \Delta \bar{s})$ comes out positive when $\Delta G \neq 0$, and
negative, except a case with $\Delta \bar{d} =0$ at $Q^2=3~(GeV/c)^2$,
for $\Delta G=0$ as it is expected to give a relevant negative
contribution to the EJ sum rule for the proton. However, with the only
exclusion of a case with $\Delta \bar{d} \neq 0$ $(\Delta s + \Delta
\bar{s})$ gives a small contribution ($-.0045$) to the EJ sum rules
and the corresponding fits practically give the same value to 
the Bj sum rule as the ones with $(\Delta s + \Delta \bar{s})=0$. 

\noindent
c) Let us consider finally the role of $\Delta G$, which gives the
most accepted solution of the {\it spin crisis} \cite{Altarelli}. We
find always positive values for $\Delta G$, which implies negative
contributions to the EJ sum rules, but its presence does not modify
the Bj sum rule, which becomes smaller at
$Q^2=3~(GeV/c)^2$, and even more at $10~(GeV/c)^2$ than the predicted
value (with QCD corrections up to the third order in $\alpha_s$). We
understand this behaviour as a result of the very strong restrictions
on our description of $\Delta G(x)$. According to Eq. (\ref{4.3}), in
fact, the function $f(x)$, which appears in the parameterization of
the gluon distributions, is determined by the unpolarized data.
Moreover, since the Bose--Einstein factor $\left[ \exp\left(x -
\tilde{x}(G)/\bar{x}\right) - 1 \right]^{-1}$ enhances the small $x$
region, $\Delta G(x)$ has a very little overlap with $\Delta u(x)$ and
$\Delta d(x)$ and consequently the value of $\Delta G$ has little
influence on $\Delta u$ and $\Delta d$. The relatively smooth
behaviour at low $x$ seen by E143, as well as the different behaviour
in the same $x$ region of $x\gpx$ and $x\gdx$ seen by SMC, hardly to
be described by a $I=0$ term as the gluon contribution, do not ask for
an important role of the anomaly for the description of the polarized
structure functions studied. 

\noindent
d) The values found for the exponents $\alpha_L$, $-1.18$ at $Q^2=3~
(GeV/c)^2$ and $-1.28$ at $Q^2=10~(GeV/c)^2$, confirm our assumption
that {\it liquid} term should not contribute to the quark parton
model sum rules, which otherwise would not converge. 

To reach an idea of the quality of the fits found, as well as the
effect of the anomaly contributions, we report in Tables II.a and II.b
the parameters and the gas abundances for partons, found with $\Delta
\bar{q}_i=0$ and with/without $\Delta G(x)$ at $Q^2 = 3$, and
$10~(GeV/c)^2$, and compare them with the results of a previous paper
\cite{bmmt}. Figures 1.a--5.a and 1.b--5.b show our theoretical
predictions versus the experimental data, for $\fnx/\fpx$, $\fpx$,
$xF_3(x)$, $x\gpx$, and $x\gdx$ corresponding to $Q^2=3$ and
$Q^2=10~(GeV/c)^2$ respectively, for the fits with $\Delta G=0$ (solid
line) and $\Delta G \neq 0$ (dashed line) reported in Tables II.a and
II.b. For the same choices of parameters, the predictions for $x\gnx$
at $Q^2=3~(GeV/c)^2$ are compared with the measurements at
$Q^2=2~(GeV/c)^2$ by E142 \cite{E142} in Figure 6., and in Figure 7.a
, 7.b the predictions for unpolarized gluon distributions are plotted
versus the experimental results \cite{gluoni1}--\cite{gluoni3}.
Note that, as already stated at the beginning of the section,
the lack of agreement at $x \lapproxeq .1$ shows the necessity
of a {\it liquid} component also for the gluons.
Finally, in Figure 8., the ratios $u^{\uparrow}(x)/d(x)$ and
$u^{\downarrow}(x)/d(x)$ are plotted for the choice of parameters of
Table II.a and II.b corresponding to $\Delta G=0$ .

\section{Conclusions}

As we expected, the narrowing of the parton distributions implied by the 
evolution equations \cite{evolv} is easily realized with a slight 
increasing of the parameters $\alpha$ and $\alpha_L$, responsible for the
low $x$ behaviour of the gas and {\it liquid} distributions respectively,
and with a slight decreasing of the {\it thermodynamical} potentials. Indeed, 
the shape-abundance correlation, so clearly indicated by data for the 
$u^{\uparrow}$ parton, may not be washed out by the logarithmic 
corrections from $Q^2=3$ to $10~(GeV/c)^2$.

The good description we get for $\fpx$, $\fnx$ and $xF_3(x)$, confirms
that the correlation shape--abundance, peculiar of our scheme based on
Pauli principle, is obeyed by the parton distributions. 

The connection between the defects in the Gottfried and Ellis--Jaffe
sum rules based on Pauli principle finds support on the fact that the
relationship 
\begin{equation}
\Delta u \simeq 2 \tilde{F} + u - d -1~~~,
\label{5.1}
\end{equation}
is rather well satisfied by the parton distributions found also with
$\Delta G \neq 0$.\\ It is also instructive to compare the behaviour
of the ratios $u^{\uparrow}(x)/d(x)$ and $u^{\downarrow}(x)/d(x)$ of
our distributions, described in Figure 8. The first ratio, in fact, is
a rapidly increasing function of $x$, whereas the second one is almost
constant $\approx .5$, in good agreement with the expected value
$1-F$. \\ The presence of a defect in the Bj sum rule also for the
fit in which we allow for 
$\Delta G \neq 0$, arises from the specific properties of our
Bose--Einstein description for gluons. In particular, the ratio
$\Delta G(x)/G(x)$ results to be large at small $x$ where the 
difference between the small negative
values of $\tilde{x}(G^{\uparrow})$ and $\tilde{x}(G^{\downarrow})$ is
relevant, and it is rapidly decreasing to about
$\mbox{tanh}(\tilde{x}(G^{\uparrow})-\tilde{x}(G^{\downarrow})/ 2
\bar{x})$ at large $x$. As a consequence of the modest overlap of
$\Delta G(x)$ and $\Delta u(x)$ and $\Delta d(x)$, $\Delta u$ and
$\Delta d$ are slightly influenced by the value of $\Delta G$. 

\vspace{.8cm}

We would like to thank Drs. Gianpiero Mangano, Ofelia Pisanti and
Pietro Santorelli for useful discussions and valuable comments.

\newpage
\bigskip\bigskip
\par\noindent
{\bf Table I.a}\\
\bigskip
{\footnotesize
\begin{tabular}{|c|c|c|c|c|c|c|}
\hline
$Q^2=3$& $\Delta \bar{u}=\Delta G=0$ 
&$\Delta \bar{u}=0$ 
&$\Delta \bar{u}=0$ 
& $\Delta(s +\bar{s})=0$
&QCD & Experimental\\
$(GeV/c)^2$&$\Delta(s +\bar{s})=0$ 
&$\Delta(s +\bar{s})=0$ 
& $\Delta G=0$ & & predictions & data \\
\hline
& & & & & &\\
$\Delta u$ & $.640$ & $.660$ & $.671$ & $.612$ & & \\
& & & & & $.782$ & \\
$\Delta \bar{u}$ & $0.$ & $0.$ & $0.$ & $.062$ & & \\
& & & & & & \\
$\Delta d$ & $-.300$ & $-.270$ & $-.263$ & $-.260$ & $-.238$ & \\
& & & & & & \\
$\Delta ( s +\bar{s})$ & $0.$ & $0.$ & $-.080$ & $0.$ &  & \\
& & & & & & \\
$\Delta G$ & $0.$ & $.40$ & $0.$ & $.94$ & & \\
& & & & & & \\
$\Gamma^p_1$ & $.126$ & $.124$ & $.130$ & $.118$ & $.161 \pm .007$
& $.127 \pm .004 \pm .010$ \cite{g1pE143}\\
& & & & & & \\
$\Gamma^n_1$ & $-.031$ & $-.031$ & $-.026$ & $-.039$ &
$-.009 \pm .005$ & $-.035 \pm .016$ \cite{g1dE143,g1pE143}\\
& & & & & & \\
$ \Gamma^p_1 - \Gamma^n_1$ & $.157$ & $.155$ & $.156$ & $.157$ & 
$.170 \pm .009$& $.163 \pm .010 \pm .016$ \cite{g1dE143}\\
& & & & & & \\
$\chi^2_{tot}/N$ & $2.33$ & $2.32$ & $2.34$ & $2.34$& & \\
& & & & & & \\
$\chi^2_{p}/N$ & $1.19$ & $1.24$ & $1.24$ & $1.25$ &  & \\
& & & & & & \\
$\chi^2_{d}/N$ & $1.16$ & $1.06$ & $1.17$ &  $1.12$& &\\
& & & & & & \\ 
\hline
\end{tabular}}
\bigskip\bigskip
\par\noindent
{\bf Table I.b}\\
\bigskip
{\footnotesize
\begin{tabular}{|c|c|c|c|c|c|c|}
\hline
$Q^2=10$& $\Delta \bar{u}=\Delta G=0$ 
&$\Delta \bar{u}=0$ 
&$\Delta \bar{u}=0$ 
& $\Delta(s +\bar{s})=0$
&QCD & Experimental\\
$(GeV/c)^2$&$\Delta(s +\bar{s})=0$ 
&$\Delta(s +\bar{s})=0$ 
& $\Delta G=0$ & & predictions & data \\
\hline
& & & & & &\\
$\Delta u$ & $.670$ & $.730$ & $.703$ & $.588$ & &\\
& & & & & $.826$ &\\
$\Delta \bar{u}$ & $0.$ & $0.$ & $0.$ & $.138$ & &\\
& & & & & &\\
$\Delta d$ & $-.290$ & $-.210$ & $-.255$ & $-.183$ & $-.268$ &\\
& & & & & &\\
$\Delta ( s +\bar{s})$ & 0. & 0. & $-.081$ & 0. & &\\
& & & & & &\\
$\Delta G$ & $0.$ & $1.5$ & $0.$ & $2.44$ & &\\
& & & & & &\\
$\Gamma^p_1$ & $.133$ & $.129$ & $.137$ & $.117$ & $.169 \pm .005$
 & $.136 \pm .011 \pm .011$ \cite{g1pSMC}\\
& & & & & &\\
$\Gamma^n_1$ & $-.027$ & $-.028$ & $-.021$ & $-.039$ &
$-.014 \pm .004$ & $-.062 \pm .029$ \cite{g1dSMC,g1pSMC}\\
& & & & & &\\
$\Gamma^p_1$ - $\Gamma^n_1$ & $.160$ & $.157$ & $.158$ & $.156$ & 
$.183 \pm .006$ & $.199 \pm .038$\cite{g1dSMC}\\
& & & & & &\\
$\chi^2_{tot}/N$ & $1.21$ & $1.17$ & $1.23$ & $1.12$ & &\\
& & & & & &\\
$\chi^2_{p}/N$ & $.60$ & $.89$ & $.67$ & $1.0$ & & \\
& & & & & &\\
$\chi^2_{d}/N$ & $2.90$ & $2.56$ & $3.22$ &  $2.62$ & &\\ 
& & & & & &\\ 
\hline
\end{tabular}}
\newpage
\bigskip\bigskip
\par\noindent
{\bf Table II.a}\\
\bigskip
\small{
\begin{tabular}{|c|c|c|c|c|c|c|}
\hline
Param.
& \multicolumn{2}{c|}{Previous fit} & 
\multicolumn{2}{c|}{Present fit with only} &
\multicolumn{2}{c|}{Present fit with only} \\
$Q^2=3$& \multicolumn{2}{c|}{\cite{bmmt}} & 
\multicolumn{2}{c|}{$\Delta u$, $\Delta d \neq 0$} &
\multicolumn{2}{c|}{$\Delta u$, $\Delta d$, $\Delta G \neq 0$}\\
$(GeV/c)^2$& \multicolumn{2}{c|}{($\chi^2/N=2.47$)} & 
\multicolumn{2}{c|}{($\chi^2/N=2.33$)} &
\multicolumn{2}{c|}{($\chi^2/N=2.32$)}\\
\hline
$A$ & \multicolumn{2}{c|}{$ 2.66\begin{array}{c} +.09\\ -.08 \end{array} $}
& \multicolumn{2}{c|}{$ 2.51\pm .07$}
& \multicolumn{2}{c|}{$ 2.54\pm .08$}\\
$\alpha$ & \multicolumn{2}{c|}{$ -.203 \pm .013$}
& \multicolumn{2}{c|}{$ -.231 \pm .012$}
& \multicolumn{2}{c|}{$ -.231 \pm .011$}\\
$\beta$ &\multicolumn{2}{c|}{$ 2.34 \begin{array}{c} +.05\\ -.06\end{array}$}
&\multicolumn{2}{c|}{$ 2.21 \pm .04$}
&\multicolumn{2}{c|}{$ 2.22 \pm .04$} \\
$A_{L}$ 
& \multicolumn{2}{c|}{$ .0895 \begin{array}{c} +.0107\\-.0084\end{array}$}
& \multicolumn{2}{c|}{$ .127 \begin{array}{c} +.016\\-.013\end{array}$}
& \multicolumn{2}{c|}{$ .128 \begin{array}{c} +.015\\-.013\end{array}$}\\
$\alpha_{L}$ & \multicolumn{2}{c|}{$ -1.19 \pm .02$}
& \multicolumn{2}{c|}{$ -1.18 \begin{array}{c} +.03\\-.02\end{array}$}
& \multicolumn{2}{c|}{$ -1.18 \pm .02$}\\
$\beta_{L}$&\multicolumn{2}{c|}{$7.66 \begin{array}{c}+1.82\\-1.59
\end{array}$}
&\multicolumn{2}{c|}{$10.3 \begin{array}{c}+1.4\\-1.3\end{array}$}    
&\multicolumn{2}{c|}{$10.1 \begin{array}{c}+1.4\\-1.3\end{array}$}\\    
\cline{2-7}
$\bar{x}$&$.235 \pm .009$ & gas abund. &$.214 \pm .008$ & gas abund.
&$.223 \pm .011$ & gas abund. \\
\cline{3-3} \cline{5-5}  \cline{7-7}
$\tilde{x}(u^{\uparrow})$& $1.00 \pm .07$ & $1.15 \pm .01$
& $1.00 \pm .02$ & $1.22 \pm .01$
& $1.00 \pm .02$ & $1.23 \pm .01$\\
$\tilde{x}(u^{\downarrow})$ & $.123 \pm .012$ & $.53 \pm .01$
& $.141 \pm .011$ & $.58 \pm .01$
& $.129 \begin{array}{c}+.014\\-.015\end{array}$ & $.57 
\begin{array}{c} +.02\\-.01\end{array}$\\
$\tilde{x}(d^{\uparrow})$& $-.068 \begin{array}{c} +.021\\
-.024\end{array}$ & $.33 \pm .03$ & $-.029 \begin{array}{c} +.019\\
-.020\end{array}$ & 
$.37 \begin{array}{c} +.02\\-.03\end{array}
$ & $-.028 \pm .020$ & $.38 \pm .03$ \\
$\tilde{x}(d^{\downarrow})$ & $.200 \begin{array}{c} +.013\\
-.014\end{array}$ & $.62 \pm .01$ & $.211 \pm .011$ & 
$.67 \pm .01$ & $.196 \begin{array}{c} +.015\\
-.016\end{array}$ & $.65 \pm .01$\\
$\tilde{x}(\bar{u}^{\uparrow})$
& $-.886 \pm .266$ & $.015 \begin{array}{c} +.034\\ -.009\end{array}$
& $-.522 \begin{array}{c}+.049\\-.061\end{array}$ 
& $.054 \begin{array}{c} +.019\\ -.017\end{array}$    
& $-.559 \begin{array}{c}+.057\\-.075\end{array}$ 
& $.052 \begin{array}{c} +.022\\-.019\end{array}$ \\
$\tilde{x}(\bar{u}^{\downarrow})$& $''$ & $''$ & $''$ & $''$
& $''$ & $''$\\
$\tilde{x}(\bar{d}^{\uparrow})$
& $-.460\begin{array}{c} +.047\\
-.064\end{array}$ &$.08 \begin{array}{c} +.03\\-.02\end{array}$
& $-.339\begin{array}{c} +.032\\
-.040\end{array}$ &$.12 \pm .04$
& $-.366\begin{array}{c} +.037\\
-.049\end{array}$ &$.12 \begin{array}{c} +.05\\-.04\end{array}$\\
$\tilde{x}(\bar{d}^{\downarrow})$ & $''$ & $''$ & $''$ & $''$ & $''$ & $''$\\
$\tilde{x}(G^{\uparrow})$
& $-.067$ & $3.16$ & $-.067\begin{array}{c} +.008 \\ -.009\end{array}$ 
& $2.9 \pm .4$
& $-.069\pm .09$ 
& $3.0 \begin{array}{c} +.05\\-.04\end{array}$\\
$\tilde{x}(G^{\downarrow})$ & $''$ & $''$ & $''$ & $''$ 
& $-.085\begin{array}{c} +.015 \\ -.019\end{array}$  
& $2.6 \begin{array}{c} +.06\\-.05\end{array}$\\
\hline
\end{tabular}}
\newpage
\bigskip\bigskip
\par\noindent
{\bf Table II.b}\\
\bigskip
\small{
\begin{tabular}{|c|c|c|c|c|}
\hline
Parameters & \multicolumn{2}{c|}{Present fit with only} &
\multicolumn{2}{c|}{Present fit with only} \\
$Q^2=10$& \multicolumn{2}{c|}{$\Delta u$, $\Delta d \neq 0$} &
\multicolumn{2}{c|}{$\Delta u$, $\Delta d$, $\Delta G \neq 0$}\\
$(GeV/c)^2$& \multicolumn{2}{c|}{($\chi^2/N=1.21$)} &
\multicolumn{2}{c|}{($\chi^2/N=1.17$)}\\
\hline
$A$ & \multicolumn{2}{c|}{$ 2.04\begin{array}{c} +.12\\ -.21 \end{array} $}
& \multicolumn{2}{c|}{$ 1.98\begin{array}{c} +.19\\ -.15 \end{array} $}\\
$\alpha$ 
& \multicolumn{2}{c|}{$ -.343 \begin{array}{c} +.023\\ -.062 \end{array} $}
& \multicolumn{2}{c|}{$ -.376 \begin{array}{c} +.046\\ -.036 \end{array}$}\\
$\beta$ 
&\multicolumn{2}{c|}{$ 2.29 \pm .04$}
&\multicolumn{2}{c|}{$ 2.28 \begin{array}{c} +.05 \\-.04 \end{array}$} \\
$A_{L}$ 
& \multicolumn{2}{c|}{$ .104 \begin{array}{c} +.027\\-.012\end{array}$}
& \multicolumn{2}{c|}{$ .108 \begin{array}{c} +.021\\-.016\end{array}$}\\
$\alpha_{L}$ 
& \multicolumn{2}{c|}{$ -1.28 \begin{array}{c} +.03\\ -.02 \end{array} $}
& \multicolumn{2}{c|}{$ -1.28 \begin{array}{c} +.03\\ -.02 \end{array}$}\\
$\beta_{L}$
&\multicolumn{2}{c|}{$7.6 \begin{array}{c}+8.0\\-1.0\end{array}$}    
&\multicolumn{2}{c|}{$10.8 \begin{array}{c}+4.7\\-4.0\end{array}$}\\    
\cline{2-5}
$\bar{x}$ 
&$.252 \begin{array}{c} +.012\\ -.035 \end{array} $ 
& gas abund.
&$.247 \begin{array}{c} +.029\\ -.021 \end{array}$ 
& gas abund. \\
\cline{3-3} \cline{5-5}  
$\tilde{x}(u^{\uparrow})$
& $1.00 \pm .01$ 
& $1.26 \pm .11$
& $1.00 \pm .01$ 
& $1.35 \begin{array}{c} +.05\\ -.07 \end{array}$\\
$\tilde{x}(u^{\downarrow})$ 
& $.102 \begin{array}{c} +.043\\ -.040 \end{array} $ 
& $.59 \begin{array}{c} +.04\\ -.01 \end{array}$ 
& $.092 \begin{array}{c} +.050\\ -.073 \end{array}$ 
& $.62 \begin{array}{c} +.03\\ -.08 \end{array}$\\
$\tilde{x}(d^{\uparrow})$ 
& $-.115 \begin{array}{c} +.101\\-.061\end{array}$ 
& $.35 \begin{array}{c} +.10\\ -.07 \end{array}$ 
& $-.051 \begin{array}{c} +.064\\-.093\end{array}$ 
& $.44 \begin{array}{c} +.06\\ -.11 \end{array}$ \\
$\tilde{x}(d^{\downarrow})$ 
& $.142 \begin{array}{c} +.047\\-.034\end{array}$ 
& $.64 \pm .04$
& $.111 \begin{array}{c} +.044\\ -.064\end{array}$ 
& $.65 \begin{array}{c} +.02\\ -.07 \end{array}$\\ 
$\tilde{x}(\bar{u}^{\uparrow})$
& $-1.0 \pm .2$ 
& $.015 \begin{array}{c} +.022\\ -.011\end{array}$    
& $-.68 \pm .18$ 
& $.053 \begin{array}{c} +.069\\-.033\end{array} $ \\
$\tilde{x}(\bar{u}^{\downarrow})$
& $''$ & $''$ & $''$ & $''$\\
$\tilde{x}(\bar{d}^{\uparrow})$
& $-.492\pm.11$ 
& $.10 \begin{array}{c} +.11\\ -.07 \end{array}$
& $-.42\pm.10$
&$.14 \begin{array}{c} +.16\\ -.09 \end{array}$\\
$\tilde{x}(\bar{d}^{\downarrow})$
& $''$ & $''$ & $''$ & $''$\\
$\tilde{x}(G^{\uparrow})$
& $-.095\begin{array}{c} +.003 \\ -.001\end{array}$ 
&$3.29 \begin{array}{c} +1.25\\ -.064 \end{array}$
& $-.076 \begin{array}{c} +.014\\-.021\end{array}$ 
&$4.3 \pm 1.0$\\
$\tilde{x}(G^{\downarrow})$& $''$ & $''$ 
& $-.121\begin{array}{c} +.033 \\ -.063\end{array}$  & 
$2.8 \pm 1.2$\\
\hline
\end{tabular}}
\newpage
{{\Large \bf Table captions}}

\bigskip

\begin{itemize}
\item[Table I.a] For four different options, the values of normalized 
$\chi^2$, the parton polarizations, and the corresponding
evaluations of $\Gamma^p_1$, $\Gamma^n_1$ and $\Gamma^p_1 -
\Gamma^n_1$ are reported, versus the QCD predictions up to $\alpha_s^3$
and the experimental data. The values correspond to $Q^2=3~(GeV/c)^2$.
\item[Table I.b] The same quantities of Table I.a, are computed for 
$Q^2=10~(GeV/c)^2$.
\item[Table II.a] The parameters and the gas abundances for partons, 
found with $\Delta \bar{q}_i=0$ and with/without $\Delta G(x)$ at 
$Q^2 = 3~(GeV/c)^2$ are reported and compared with the results of 
Ref. \cite{bmmt}. Note that, no antiquarks or strange quark polarization
is assumed.
\item[Table II.b] The same quantities of Table II.a are shown for 
$Q^2 = 10~(GeV/c)^2$.
\end{itemize}

\bigskip

{{\Large \bf Figure captions}}

\bigskip

\begin{itemize}
\item[Figure 1.a] The prediction for $\fnx/\fpx$ at $Q^2=3~(GeV/c)^2$
is plotted and compared with the experimental data \cite{f2p-3g-2,f2n/f2p},
the solid line and the dashed line corresponds to the fit with $\Delta G=0$ 
and $\Delta G\neq0$ of Table II.a, respectively. This notation is valid
for all Figures 1.a-5.a, 6., and 7.a .
\item[Figure 1.b] The same quantity of Figure 1.a is plotted for 
$Q^2=10~(GeV/c)^2$,
the solid line and the dashed line corresponds to the fit with $\Delta G=0$ 
and $\Delta G\neq0$ of Table II.b, respectively. This notation is valid
for all Figures 1.b-5.b, and 7.b .
\item[Figure 2.a] The prediction for $\fpx$ at $Q^2=3~(GeV/c)^2$
is plotted and compared with 
the experimental data \cite{f2p-3g-1}, \cite{f2p-3g-2} and \cite{f2p-3g-3}.
\item[Figure 2.b] The same quantity of Figure 2.a is plotted for 
$Q^2=10~(GeV/c)^2$.
\item[Figure 3.a] $xF_{3}(x)$ is plotted for $Q^2=3~(GeV/c)^2$
and the experimental values are taken from \cite{xf3-3g}.
\item[Figure 3.b] $xF_{3}(x)$ is plotted for $Q^2=10~(GeV/c)^2$
and the experimental values are taken from \cite{xf3-3g}.
\item[Figure 4.a] $x\gpx$ at $Q^2=3~(GeV/c)^2$ is plotted and 
compared with the data \cite{g1pE143}.
\item[Figure 4.b] The same quantity of Figure 4.a corresponding
to $Q^2=10~(GeV/c)^2$ is plotted versus the experimental data \cite{g1pSMC}.
\item[Figure 5.a] $x\gdx$ at $Q^2=3~(GeV/c)^2$ is plotted and 
compared with the data \cite{g1dE143}.
\item[Figure 5.b] The same quantity of Figure 5.a corresponding
to $Q^2=10~(GeV/c)^2$ is plotted versus the experimental data \cite{g1dSMC}.
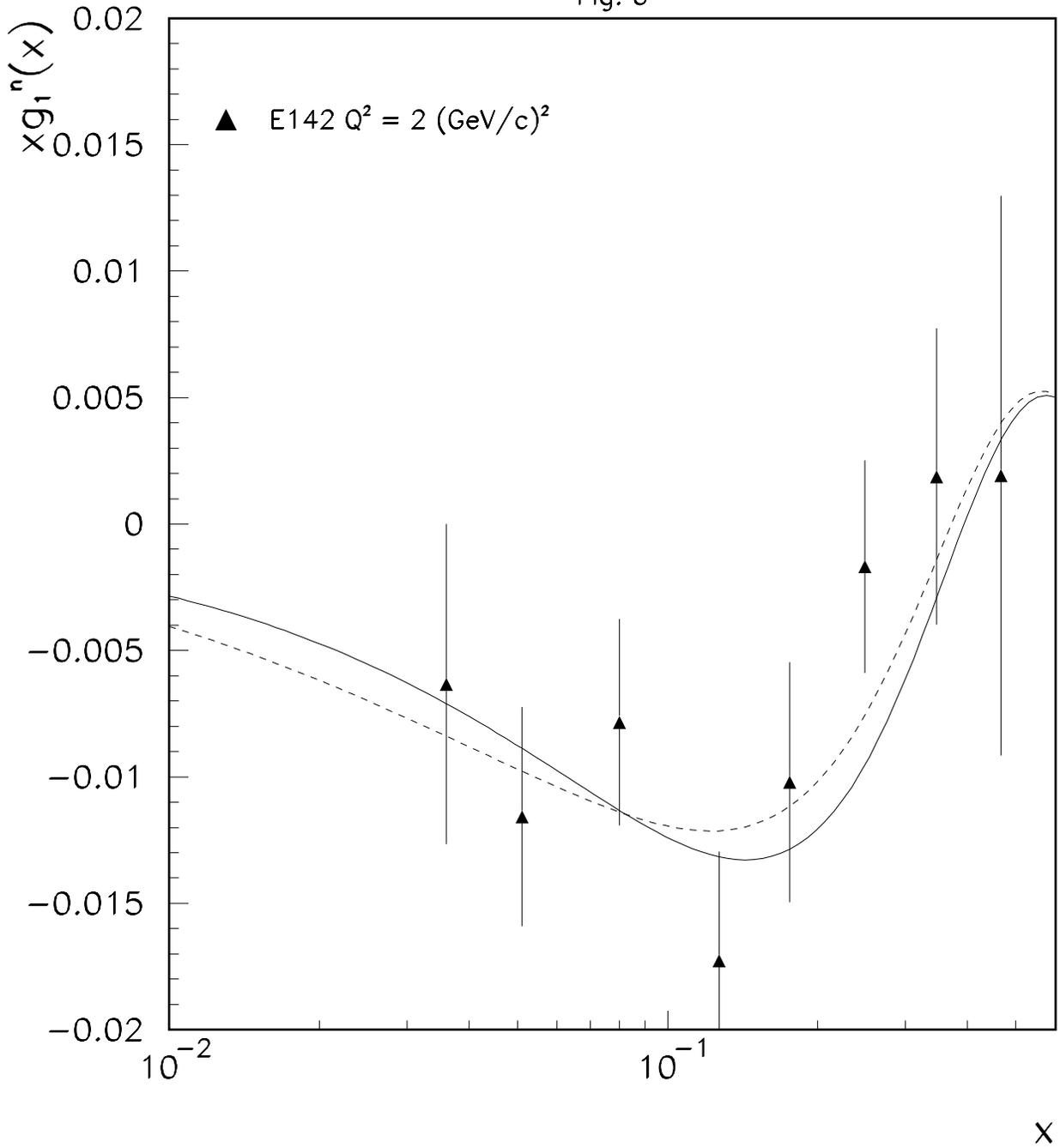
\item[Figure 6.] $x\gnx$ at $Q^2=3~(GeV/c)^2$ is plotted versus
the experimental data at $Q^2=2~(GeV/c)^2$ by E142 \cite{E142}.
\item[Figure 7.a] The predictions for unpolarized gluon distribution
are $Q^2=3~(GeV/c)^2$ are reported versus the experimental results 
\cite{gluoni1}--\cite{gluoni3}.
\item[Figure 7.b] The same quantity of Figure 7.a, but corresponding to
$Q^2=10~(GeV/c)^2$ is shown.
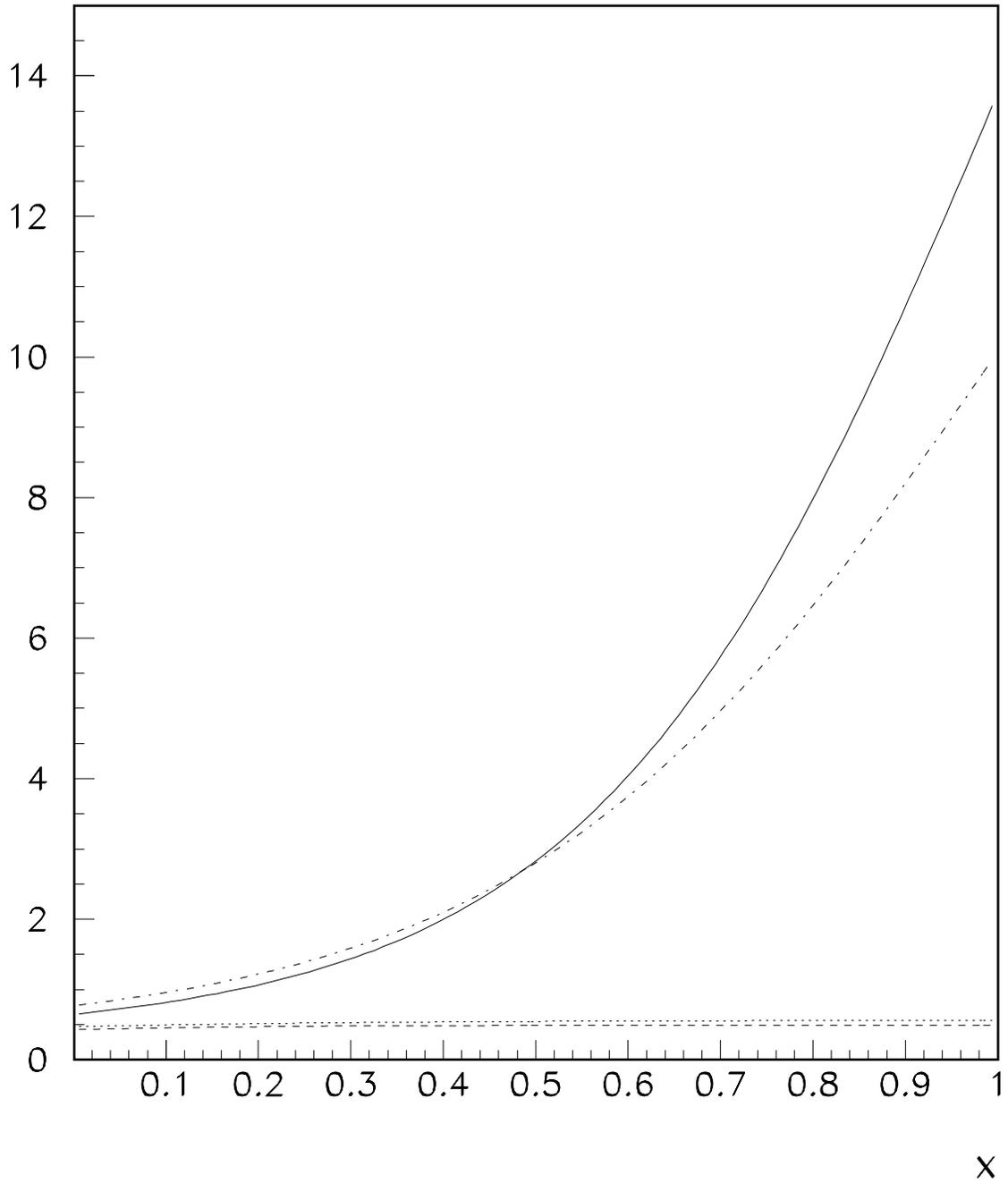
\item[Figure 8.] The solid line represents the ratio $u^{\uparrow}(x)/d(x)$,
and the dashed one $u^{\downarrow}(x)/d(x)$, for the choice of parameters 
corresponding to the second column of Table II.a. The dashed-dotted line
and the dotted one, represent the same quantities but corresponding
to the first column of Table II.b .
\end{itemize}
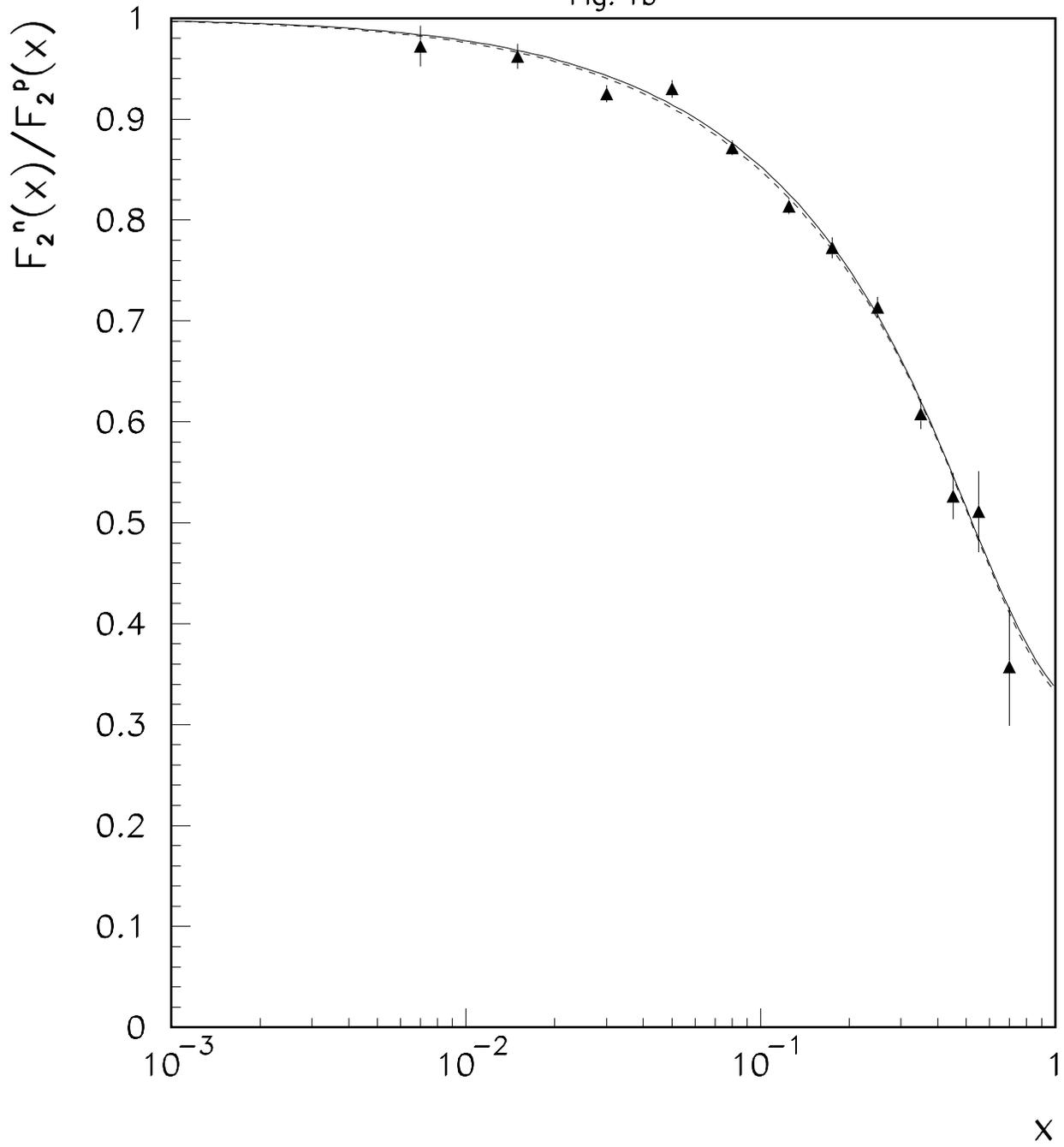
\end{document}